\documentstyle[prl,floats,aps,twocolumn,epsf,graphicx]{revtex}

\begin{document}
\twocolumn[\hsize\textwidth\columnwidth\hsize\csname
@twocolumnfalse\endcsname

\title{Wave Propagation in Nontrivial Backgrounds}
\author{Shahar Hod}
\address{The Racah Institute of Physics, The
Hebrew University, Jerusalem 91904, Israel}
\date{\today}

\maketitle

\begin{abstract}

\ \ \ It is well known that waves propagating in a nontrivial 
medium develop ``tails''. However, the
exact form of the late-time tail has so far been determined only for a
narrow class of models. We present a systematic analysis of the tail
phenomenon for waves propagating under the influence of a {\it general}
scattering potential $V(x)$. It is shown that, generically, the
late-time tail is determined by spatial {\it derivatives} of the potential. 
The central role played by derivatives of the scattering potential 
appears not to be widely recognized.
The analytical results are confirmed by numerical calculations.

\end{abstract}
\bigskip

]

One of the most remarkable features of wave dynamics in curved
spacetimes is the development of ``tails''. Gravitational waves (or
other fields) propagate not only along light cones, but also spread
inside them. This implies that at late times waves do not cut
off sharply but rather die off in tails. 
In particular, it is well established that the late-time evolution of
massless fields propagating in black-hole spacetimes is dominated by 
an inverse {\it power-law} behaviour.
 
Price \cite{Price} was the first to provide a detailed analysis of 
the mechanism by which the
spacetime outside of a (nearly spherical) collapsing star divests itself of
all radiative multipole moments, and leaves behind a Schwarzschild
black hole. The evolution of such waves (gravitational,
electromagnetic, and scalar) 
in a curved spacetime is governed by the Klein-Gordon (KG) equation \cite{Chan}

\begin{equation}\label{Eq1}
\left[ {{\partial ^2} \over {\partial t^2}}-{{\partial ^2} \over
    {\partial x^2}} +{{l(l+1)} \over {x^2}} +{1 \over {x^2_s}}V(x/x_s) \right]\Psi^l=0\  ,
\end{equation}
where the term $l(l+1)/x^2$ is the well-known centrifugal barrier 
($l$ is the multipole order of the field), and $V(x)$ is an effective
curvature potential (we henceforth take $x_s=1$ without loss of
generality). It was demonstrated that all radiative perturbations decay
asymptotically as an inverse power of time, 
the power indices equal $2l+3$. 
Physically, these inverse
power-law tails are associated with the backscattering of waves off
the effective curvature potential, $V(x)$, at asymptotically far regions
\cite{Thorne,Price}.

The analysis of Price has been extended by many authors. Bi\v{c}\'{a}k
\cite{Bicak} generalized the results to the case of an
electrically neutral scalar field propagating 
in a charged Reissner-Nordstr\"om spacetime. 
Leaver \cite{Leaver} demonstrated that the late-time tail can be
associated with the existence of a branch cut in the Green's function
for the wave propagation problem. 

Gundlach, Price, and Pullin \cite{Gundlach2} showed that 
power-law tails are a genuine feature of gravitational collapse --
the existence of these tails was demonstrated in full {\it non}-linear 
numerical simulations of the spherically symmetric collapse 
of a self-gravitating scalar 
field (this was later reproduced in \cite{BurOr}). Moreover, since the
late-time tail is a direct consequence of the scattering of the waves at 
asymptotically far regions, it has been suggested that power-law tails
would develop independently of the existence of an 
horizon \cite{Gundlach1}. This implies that tails should appear in
perturbations of stars as well. 
Moreover, the KG wave equation has a much
wider range of physical applicability. For instance, electromagnetic waves
propagating in an optical cavity can be described by this equation as
well \cite{Lang,Ching} [A nontrivial dielectric constant
distribution, $n^2(x)$, provides a nontrivial medium, and 
is analogous to the presence of a scattering potential $V(x)$.] 
For other related
works, see e.g.,
\cite{Ching,SuPr,Andersson1,Andersson2,Brad1,Brad2,HodPir123,HodPir4,Barack1,Hod1,KokSch},
and references therein. 

The dynamics of waves in {\it rotating} black-hole spacetimes has
received much attention recently 
\cite{Krivan1,Krivan2,Ori1,Barack2,Hod2,BarOr,Hod3,Hod4,Barack3,Krivan3,AnderKos}.
It has been demonstrated that power-law tails present in the Kerr
spacetime as well (However, the damping indices are generally
different from those found in the Schwarzschild case.)

Yet, a thorough understanding of the fascinating phenomenon of wave
tails is far from being complete. In particular, the exact form of the late-time tail
has so far been determined only for a narrow class of models. In fact, most
of previous analyses are restricted to ``logarithmic
potentials'' of the form $V(x) \sim \ln^{\beta}x/x^{\alpha}$ 
(where $\alpha >2$ and $\beta=0,1$ are parameters). In a brilliant
work, Ching et. al. \cite{Ching} have
shown that,  generically, the late-time behaviour of waves propagating under the
influence of these specific potentials has the form $\Psi \sim
\ln^{\beta}t/t^{2l+\alpha}$ (with the well-known exception of a ``pure''
power-law decay in the Schwarzschild spacetime).

The purpose of this paper is to present a systematic analysis of the
tail phenomenon with a {\it general} scattering potential $V(x)$. 
Ching et. al. \cite{Ching} provided a heuristic picture of the
scattering problem which is an important first step in this
direction: Consider a wave from a source point $y$. 
The late-time tail observed at a fixed spatial location, $x$, is a
consequence of the wave first propagating to a 
distant point $x' >>y,x$, being scattered by $V(x')$, and then returning
to $x$ at a time $t \simeq (x'-y) + (x'-x) \simeq 2x'$. Hence, 
according to this scenario, the
scattering amplitude (and thus the late-time tail itself) are expected
to be proportional to $V(x') \simeq V(t/2)$.

However, it is well-known, at least in the specific case of logarithmic
potentials, that this simple picture requires two 
modifications. First, there is an {\it extra suppression} of the late-time
tail by a factor of $t^{-2l}$. Second, if $\alpha$ is an odd integer less than $2l+3$, the
leading term in the late-time tail vanishes \cite{Ching}, and one should consider
sub-leading terms (The well-known Schwarzschild spacetime belongs to
this case.)

There are several interesting and important open questions regarding
the tail phenomenon in a {\it general} analysis. 
What determines the late-time tail -- is
it simply the asymptotic form of the scattering potential itself, as suggested
by the heuristic picture ? How generic is the suppression of waves 
with $l \geq 1$ ? Is it always by a factor of $t^{-2l}$ ? What is the
(most) general class of scattering potentials for which the leading
term in the late-time tail vanishes ? (The Schwarzschild spacetime is
only one specific example to these potentials.) What is the
general form of the late-time behaviour in these cases ? 
These questions, and several others 
call for a study of the {\it general} properties of wave tails. 
In this paper we present our main results for this fascinating phenomena. 

We consider the evolution of a wave field whose dynamics is governed 
by a KG-type equation $\Phi_{;\nu}^{;\nu} +V(x)\Phi=0$. 
Resolving the field into spherical harmonics 
$\Phi =\sum\limits_{l,m} \Psi^l_m(t,r) Y^m_l(\theta,\varphi)/r$ ($r$
being the circumferential radius), 
one obtains a wave equation
of the form Eq. (\ref{Eq1}) for each
multipole moment \cite{Note} 
(For brevity we henceforth suppress the indices $l,m$ on $\Psi$.)

It proofs useful to introduce the double-null coordinates $u\equiv t-x$ 
and $v \equiv t+x$, which are a retarded time coordinate and an advanced
time coordinate, respectively. 
The initial data is in the form of some compact outgoing
pulse in the range $u_0 \leq u \leq u_1$, 
specified on an ingoing null surface $v=v_0$.

The general solution to the wave-equation (\ref{Eq1}) can be written
as a series depending on two arbitrary functions $F$ and $G$ \cite{Price}

\begin{eqnarray}\label{Eq2}
\Psi& = & \sum\limits_{k=0}^l A_k^l x^{-k} \left[ {{G^{(l-k)}(u)+(-1)^k 
F^{(l-k)}(v)}} \right]\nonumber \\
 &&+  \sum\limits_{k=0}^\infty  {B_k^l(x)
\left[ G^{(l-k-1)}(u)+({-1})^kF^{(l-k-1)}(v) \right]}\  ,
\end{eqnarray}
where $A_k^l=(l+k)!/2^kk!(l-k)!$. 
For any function $H$, $H^{(k)}$ is its $k$th
derivative; negative-order derivatives are to be interpreted as
integrals. The functions $B_k^l(x)$ satisfy the recursion relation 

\begin{equation}\label{Eq3}
B_k'={1 \over 2}\left[B''_{k-1}-l(l+1)x^{-2}B_{k-1}
  -V(A_kx^{-k} +B_{k-1}) \right]\  ,
\end{equation}
for $k \geq 1$, 
where $B'\equiv dB/dx$, and $B'_{0}=-V(x)/2$. 

For the first Born approximation to be valid the scattering potential
$V(x)$ should approach zero faster than $1/x^2$ as $x \to
\infty$, see e.g., \cite{Ching,HodPir123}. 
Otherwise, the scattering potential cannot be neglected at
asymptotically far regions [see Eq. (\ref{Eq5}) below].
It is useful to classify the
scattering potentials into three groups, according to their asymptotic
behaviour: 

\begin{itemize}
\item Group I: $|V'|$ approaches
  zero {\it slower} than $|V|/x$, and {\it faster} than $|V|$ 
as $x \to \infty$.
\item Group II: $|V'|$ approaches zero at the {\it same} rate as $|V|$ as $x \to \infty$.
\item Group III: $|V'|$ approaches zero at the {\it same} 
rate as $|V|/x$ as $x \to \infty$.

\end{itemize}

{\it Group I}. --- 
For this case the recursion relation, Eq. (\ref{Eq3}), yields
$B_k(x)=-2^{-(k+1)} V^{(k-1)}(x)$.

The first stage of the evolution is the scattering
of the field in the region $u_0 \leq u \leq u_1$. 
The first sum in Eq. (\ref{Eq2}) represents the primary waves in the
wave front (i.e., the zeroth-order solution, with $V \equiv 0$), 
while the second sum represents backscattered waves. The interpretation of
these integral terms as backscatter comes from the fact that they depend on
data spread out over a {\it section} of the past light cone, while outgoing
waves depend only on data at a fixed $u$ \cite{Price}.

After the passage of the primary waves there is no outgoing radiation for $u >
u_1$, aside from backscattered waves. This means that $G(u_1) = 0$. Hence, for a
large $x$ at $u = u_1$, 
the dominant term in Eq. (\ref{Eq2}) is

\begin{equation}\label{Eq4}
\Psi(u=u_1,x)=B_l(x)G^{(-1)}(u_1)\  .
\end{equation}
This is the dominant backscatter of the primary waves.

With this specification of characteristic data on $u=u_1$, we shall
next consider the asymptotic evolution of the field. We confine our
attention to the region $u>u_1$, $x \gg x_s$. 
To a {\it first} Born approximation, the spacetime in this region is
approximated as flat \cite{Price,Gundlach1}. 
Thus, to first order in $V$ (that is, in a first Born approximation) 
the solution for $\Psi$ can be written as

\begin{equation}\label{Eq5}
\Psi =  \sum\limits_{k=0}^l A_k x^{-k} \left[ {{g^{(l-k)}(u)+(-1)^k 
f^{(l-k)}(v)}} \right]\  .
\end{equation}
Comparing Eq. (\ref{Eq5}) with the initial data on $u=u_1$, Eq. (\ref{Eq4}), one
finds $f(v)=-2^{-1} V^{(-1)}(v/2) G^{(-1)}(u_1)$.

For late times $t \gg x$ one can expand
$g(u)=\sum\limits_{n=0}^{\infty} (-1)^ng^{(n)}(t)x^n/n!$ and similarly
for $f(v)$. With these expansions, Eq. (\ref{Eq5}) can be rewritten as

\begin{equation}\label{Eq6}
\Psi =\sum\limits_{n=-l}^{\infty}  {K_l^nx^n\left[ f^{(l+n)}(t)+(-1)^ng^{(l+n)}(t) \right]}\  ,
\end{equation}
where the coefficients $K_l^n$ are those given in \cite{Price}. They
vanish for $-l \leq n \leq l$ if $l-n$ is odd.

Using the boundary conditions for small $r$ (regularity as $x \to -\infty$, at
the horizon of a black hole, or at $x=0$, for a stellar model), one
finds that at late times $g(t)=(-1)^{l+1}f(t)$ to first order in the 
scattering potential $V$ (see e.g., \cite{Price,Gundlach1} for
additional details). 
That is, the incoming and outgoing parts of
the tail are equal in magnitude at late-times. This almost total
reflection of the ingoing waves at small $r$ can easily be understood
on physical grounds -- it 
simply manifests the impenetrability of the barrier to low-frequency
waves \cite{Price} (which are the ones to dominate the late-time 
evolution \cite{Leaver}). 
We therefore find that the late-time behaviour 
of the field at a fixed radius ($x \ll t$) 
is dominated by [see Eq. (\ref{Eq6})]

\begin{equation}\label{Eq7}
\Psi \simeq 2K_l^{l+1}f^{(2l+1)}(t)x^{l+1}\  , 
\end{equation}
which implies

\begin{equation}\label{Eq8}
\Psi \simeq \Psi_0 G^{(-1)}(u_1)x^{l+1} V^{(2l)}(t/2)\  ,
\end{equation}
where $\Psi_0=-2^{-(2l+1)}K_l^{l+1}$. Hence, the late-time tail is
determined by the $2l$th {\it derivative} of the scattering potential.

The analysis for Groups II and III proceeds along the same lines. That
is, one should first solve Eq. (\ref{Eq3}) for $B_k(x) \to$ find
$f(v)$ using Eq. (\ref{Eq5}) $\to$ which finally yields $\Psi(t \to \infty)$ through
Eq. (\ref{Eq6}). In the following we present the main results for
groups I and II.

{\it Group II}. --- 
The dominant backscatter of the primary waves is 
$\Psi(u=u_1,x)=\sum\limits_{k=l}^{\infty} 
B_k(x)G^{(l-k-1)}(u_1)$, where the $B_k$ are the same as for group I. 
Using an analysis along the same lines as before, one finds 

\begin{eqnarray}\label{Eq9}
\Psi& \simeq &
-\sum\limits_{n=l+1,l+3,...}^{\infty} 2^{-n}K_l^n x^n
\sum\limits_{k=l}^{\infty} 2^{-k}\nonumber \\ 
&& \times G^{(l-k-1)}(u_1)V^{(n+k-1)}(t/2)\  ,
\end{eqnarray}
at late-times. Note that Eq. (\ref{Eq9}) is merely a generalization of
Eq. (\ref{Eq8}), and reduces to it if $|V'|$ approaches zero faster
than $|V|$ (in which case $V^{(2l)}$ dominates at late-times).

{\it Group III}. --- 
Let $V(x) \equiv W(x)/x^{\alpha}$, 
where $\alpha >2$, and $W(x)/x^{\gamma} \to 0$ as $x \to \infty$ for
any $\gamma>0$. The {\it general} solution for $B_k(x)$ is now
given by $B_k(x) \simeq \sum\limits_{n=0}^{\infty} 
b_n(k,\alpha)W^{(n)}(x)/x^{\alpha+k-n-1}$, 
where the $b_n(k,\alpha)$ are complicated numerical coefficients, which are 
determined by the recursion relation Eq. (\ref{Eq3}). 
The analysis now proceeds along the same lines as before; One finds that 
the late-time behaviour of the wave is given by 

\begin{equation}\label{Eq10}
\Psi \simeq G^{(-1)}(u_1)x^{l+1} 
\sum\limits_{n=0}^{\infty} c_n(l,\alpha)W^{(n)}(t/2)t^{-(\alpha+2l-n)}\  ,
\end{equation}
where the coefficients $c_n(l,\alpha)$ are constructed 
from the $b_n(l,\alpha)$. 

Group III is divided
into three subgroups according to the asymptotic behaviour of the
function $W(x)$, and the value of $\alpha$:

Subgroup IIIa: $|W'|$ approaches zero {\it faster} than $|W|/x$ 
as $x \to \infty$, and $\alpha$ is {\it not} an odd integer less than
$2l+3$. In this case, the dominant term in Eq. (\ref{Eq10}) 
is $W(t/2)t^{-(\alpha+2l)}$. Recall now that $V' \sim W/t^{\alpha+1}$ as
$t \to \infty$, which implies 

\begin{equation}\label{Eq11}
\Psi \sim  G^{(-1)}(u_1)x^{l+1} V^{(2l)}(t/2)\  ,
\end{equation}
at late-times.

Subgroup IIIb: $|W'|$ approaches zero {\it faster} than $|W|/x$ 
as $x \to \infty$, and $\alpha$ is an odd integer less than
$2l+3$. (This subgroup of scattering 
potentials includes the Schwarzschild spacetime as a special case.) 
In this case one finds that the leading term of $B_l(x)$
[proportional to $V^{(l-1)}(x)$] {\it vanishes}, and 
sub-leading terms should therefore be considered. Hence, 
the late-time behaviour of the wave is dominated by 

\begin{equation}\label{Eq12}
\Psi \sim G^{(-1)}(u_1) x^{l+1} V^{(2l)}_{sl}(t/2)\  .
\end{equation}
Here $V^{(2l)}_{sl}$ is the (first) {\it sub-leading} term in
the asymptotic derivative. Namely, $V^{(2l)}_{sl} \sim
W'/t^{\alpha+2l-1}$. [Note that the results of \cite{Ching} for the specific family of 
logarithmic potentials (with $W \sim {\ln}^{\beta} x$), 
coincide with the general expressions, Eqs. (\ref{Eq11}) and (\ref{Eq12}).]

Subgroup IIIc: $|W'|$ approaches zero at the {\it same} rate as $|W|/x$
  as $x \to \infty$. In this case, the late-time dynamics of the field
  is given by Eq. (\ref{Eq10}).

{\it Numerical calculations}. --- 
It is straightforward to integrate Eq. (\ref{Eq1}) using the methods 
described in \cite{Gundlach1,Hod1}. 
The {\it late-time} evolution of the 
field is independent of the form of the initial data used. The
results presented here are for a Gaussian pulse. 

Table \ref{Tab1} gives a selected list
of scattering potentials, chosen as representative of the various different
groups (We have studied other potentials as well, which are not shown
here.) 

\begin{table}
\caption{Late-time behaviour for various scattering potentials.}
\label{Tab1}
\begin{tabular}{lll}
Group &$V(x)$ & $\Psi^l(t \to \infty)$\\
\tableline
I & $e^{-x^{\beta}}, \  0<\beta<1$ & $e^{-(t/2)^{\beta}}t^{-2l(1-\beta)}$ \\
I & $\sin(x^{\beta})/x^{\alpha},\  0<\beta<1$ & $\sin[(t/2)^{\beta}]t^{-\alpha-2
l(1-\beta)} $ \\
II & $\sin(x)/x^{\alpha}$ & periodic fun. $\times t^{-\alpha} $ \\
IIIa & $x^{1/x^{\beta}}/x^{\alpha}, \ \beta >0$ & $t^{-(\alpha+2l)}$  \\
IIIb & $x^{1/x^{\beta}}/x^{\alpha}, \ \alpha$ odd $< 2l+3, \ \beta >0$ &
$t^{-(\alpha+2l+\beta)}\ln t$  \\
\end{tabular}
\end{table}

\begin{figure}[tbh]
\centerline{\epsfxsize=9cm \epsfbox{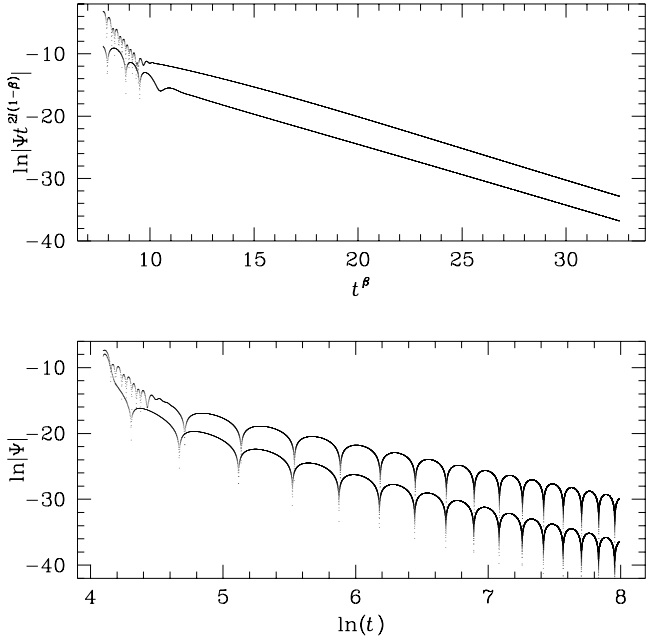}} 
\caption{Temporal evolution of the field for scattering potentials which
belong to Group I. Top panel: $V(x)=e^{-(x/x_0)^{\beta}}$ 
(the results presented here are
for the parameters $\beta=1/2$ and $x_0=1/2$). 
We display the quantity $\ln |\Psi t^{2l(1-\beta)}|$ vs
$t^{\beta}$. There is a definite 
linear dependence at late-times, with slopes of $-0.99$
and $-1.02$ for $l=0$ (lower graph) and $l=1$, respectively. These are in excellent
agreement with the analytically predicted value of $-1$. 
Bottom panel: $V(x)= \sin[(x/x_0)^{\beta}]/x^{\alpha}$ (the results presented here are
for the parameters $\alpha=3$, $\beta=1/2$, and $x_0=1/2$). 
An oscillatory power-law fall off is manifest at late times. 
The power-law indices (determined from the maxima of the oscillations)
are $-4.05,$ and $-5.07$, for $l=1$ (upper graph), and $l=2$,
respectively. These are in excellent agreement
with the {\it analytically} predicted values of $-4,$ and $-5$.}
\label{Fig1}
\end{figure}

\begin{figure}[tbh]
\centerline{\epsfxsize=9cm \epsfbox{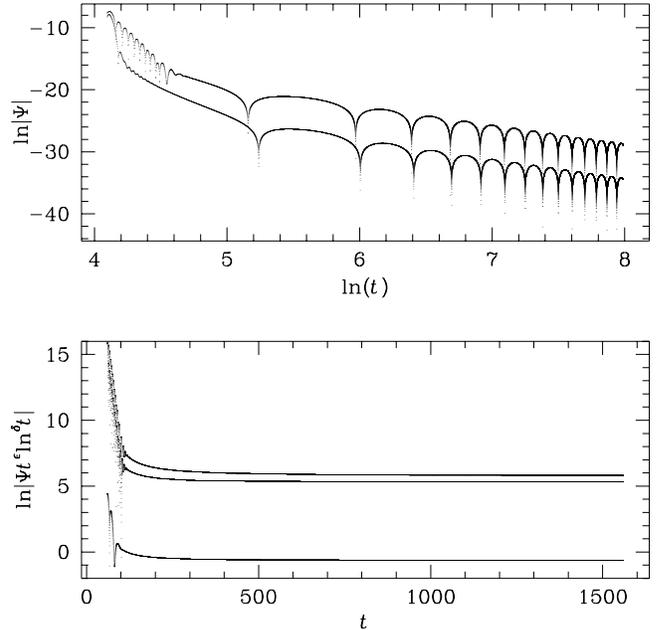}} 
\caption{Top panel: Temporal evolution of the field for a scattering 
potential of the form 
$V(x)= \sin(x/x_0)/x^{\alpha}$, which belongs to Group II 
(the results presented here are
for the parameters $\alpha=3$ and $x_0=100/\pi$). 
The power-law indices 
are $-3.02$ and $-3.04$ for $l=1$ (upper graph) and $l=2$, respectively. These
values are in excellent agreement
with the {\it analytically} predicted value of $-3$ for {\it both} $l=1$ and
$l=2$. The oscillations period agrees with the predicted value 
to within $0.4\%$. 
Bottom panel: Temporal evolution of the field for scattering potentials of the form 
$V(x)= x^{1/x^{\beta}}/x^{\alpha}$, which belong to Group III 
(the results presented here are for the case $\beta=1$). 
We present the quantity $\ln |\Psi t^{\epsilon} \ln ^{\delta} t|$ vs $t$, where:
(i) $\epsilon=3$, $\delta=0$ for $\alpha=3$ with $l=0$,
(ii) $\epsilon=6$, $\delta=-1$ for $\alpha=3$ with $l=1$, and 
(iii) $\epsilon=6$, $\delta=0$ for $\alpha=4$ with $l=1$ 
(from bottom to top). These
quantities are analytically predicted to approach a {\it constant}
value at asymptotically late-times.}
\label{Fig2}
\end{figure}

The temporal evolutions of the waves 
(under the influence of the various scattering
potentials) are shown in Figs. \ref{Fig1} and \ref{Fig2}. 
We find an excellent agreement between the {\it analytical} results 
and the {\it numerical} calculations. 

{\it Summary and physical implications}. --- 
We have given a systematic analysis of the tail phenomena for 
waves propagating under the influence of 
a {\it general} scattering potential. 

It was shown that the late-time tail is governed by
spatial {\it derivatives} of the scattering potential (generically, by
the $2l$th derivative). In particular, 
the potential function itself does not enter into the expression of the
late-time tail (with the exception of the monopole case). 
The central role played by derivatives of the scattering potential 
appears not to be widely recognized. The analytical results are in
excellent agreement with numerical calculation.

In addition, we have demonstrated 
that the (extra) suppression of waves by a factor of
$t^{-2l}$ (which adds to the basic late-time decay), 
a phenomena well-known in black-hole spacetimes, 
is actually {\it not} a generic feature of the scattering
problem. In particular, for scattering potentials that belong to group
I the suppression of the waves is {\it smaller}, while for scattering potentials 
that belong to group II there is {\it no} (extra) suppression at all. 

Moreover, it was shown that the familiar case of the
Schwarzschild spacetime belongs to 
a wider group of scattering potentials, in 
which the leading term in the tail 
[proportional to $V^{(2l)}(t/2)$] {\it vanishes} (and thus, 
{\it sub-leading} terms dominate the late-time dynamics).

We are at present extending the analysis to 
include: (i) time-dependent scattering potentials, and (ii) scattering
potentials that lack spherical symmetry (in which case the scattering
problem is of $2+1$ dimensions).

\bigskip
\noindent
{\bf ACKNOWLEDGMENTS}
\bigskip

I thank Tsvi Piran for discussions. 
This research was supported by a grant from the Israel Science Foundation.


\begin{thebibliography}{99}
\bibitem{Note} For a non-singular spacetime, e.g., that 
of a star, $r \in (0,\infty)$ maps into $x \in (0,\infty)$, while for
a spacetime with an event horizon, $r$ maps 
into $x \in (-\infty,\infty)$ 
($x$ being the so-called ``tortoise'' coordinate) 
\cite{Price,Ching}.
\end{thebibliography}
\end{document}